\documentclass[11pt,a4paper]{article}
\usepackage{txfonts} 
\usepackage{upgreek}
\usepackage{graphicx}
\usepackage{subfigure}
\usepackage[left=2.50cm, right=2.50cm, top=2.00cm, bottom=2.00cm]{geometry}
\usepackage[utf8]{inputenc}
\usepackage{authblk}

\title{The Link Between the Local Bubble and Radioisotopic Signatures on Earth}

\author[1,2]{Jenny Feige\thanks{feige@astro.physik.tu-berlin.de}}
\author[1]{Dieter Breitschwerdt}
\author[3]{Anton Wallner}
\author[1]{Michael M.\ Schulreich}
\author[4]{Norikazu Kinoshita}
\author[5]{Michael Paul}
\author[6]{Christian Dettbarn}
\author[3]{L.\ Keith Fifield}
\author[2]{Robin Golser}
\author[7]{Maki Honda}
\author[8]{Ulf Linnemann}
\author[9]{Hiroyuki Matsuzaki}
\author[10]{Silke Merchel}
\author[10]{Georg Rugel}
\author[2]{Peter Steier}
\author[3]{Stephen G.\ Tims}
\author[2]{Stephan R.\ Winkler}
\author[11]{Takeyasu Yamagata}
\affil[1]{Department of Astronomy and Astrophysics, Berlin Institute of Technology, Hardenbergstr. 36, 10623 Berlin, Germany}
\affil[2]{University of Vienna, Faculty of Physics - Isotope Research and Nuclear Physics, VERA Laboratory, W\"ahringerstr. 17, 1090 Vienna, Austria}
\affil[3]{Department of Nuclear Physics, The Australian National University, Canberra, ACT 2601, Australia}
\affil[4]{Institute of Technology, Shimizu Corporation, Tokyo 135-8530, Japan}
\affil[5]{Racah Institute of Physics, The Hebrew University of Jerusalem, Jerusalem 91904, Israel}
\affil[6]{Astronomisches Rechen-Institut, Zentrum f\"ur Astronomie der Universit\"at Heidelberg, M\"onchhofstra\ss e 12-14, 69120 Heidelberg, Germany}
\affil[7]{Graduate School of Pure and Applied Sciences, University of Tsukuba, Ibaraki 305-8577, Japan}
\affil[8]{Senckenberg Collections of Natural History Dresden, GeoPlasmaLab, K\"onigsbr\"ucker Landstra\ss e 159, 01109 Dresden, Germany}
\affil[9]{MALT (Micro Analysis Laboratory, Tandem Accelerator), The University Museum, The University of Tokyo, Tokyo 113-0032, Japan}
\affil[10]{Helmholtz-Zentrum Dresden-Rossendorf, Bautzner Landstr. 400, 01328 Dresden, Germany}
\affil[11]{Graduate School of Integrated Basic Sciences, Nihon University, Tokyo 156-8550, Japan}

\date{Received: August 20, 2016}

\begin{document}
\maketitle

\begin{abstract}
\noindent
Traces of 2-3\,Myr old $^{60}$Fe were recently discovered in a manganese crust and in lunar samples. We have found that this signal is extended in time and is present in globally distributed deep-sea archives. A second 6.5-8.7\,Myr old signature was revealed in a manganese crust. The existence of the Local Bubble hints to a recent nearby supernova-activity starting 13\,Myr ago. With analytical and numerical models generating the Local Bubble, we explain the younger $^{60}$Fe-signature and thus link the evolution of the solar neighborhood to terrestrial anomalies.
\end{abstract}



\section{Introduction}

Massive stars produce and eject long-lived radionuclides -- unstable isotopes with half-lives in the order of million years or higher. Amongst these $^{26}$Al (t$_{1/2}$=0.7\,Myr) and $^{60}$Fe (t$_{1/2}$=2.6\,Myr) are of particular interest as they have been detected directly in the interstellar medium via $\upgamma$-ray observations pointing to recent nucleosynthesis in our galaxy \cite{diehl16}. Live $^{60}$Fe was also discovered in deep-sea ferromanganese crusts from the Pacific Ocean \cite{knie99,knie04}, very slow-growing (mm~Myr$^{-1}$) archives that maintain time information. A significant $^{60}$Fe-signal was detected \cite{knie04} in a layer containing the time-period of 1.7-2.6\,Myr 
which was later confirmed in the same crust \cite{fitoussi08}. The primordial $^{60}$Fe nuclei have decayed long ago and there are no natural sources producing $^{60}$Fe in-situ on Earth. Its detection can be associated with recent cosmic events, such as nearby supernova explosions. These events should leave a global trace. A study of a deep-sea sediment from the Atlantic Ocean, however, showed no evidence of an intense signal comparable to the $^{60}$Fe-peak in the Pacific ferromanganese crust \cite{fitoussi08}. The sediment accumulation rate of 3\,cm~kyr$^{-1}$ was four orders of magnitude higher than the growth rate of the crust, leading to a much better time resolution. But the proportionately higher amount of stable iron might have diluted the concentration of $^{60}$Fe to amounts close to the detection limit. Thus, the questions whether $^{60}$Fe was distributed globally and how long the Earth was subject to the $^{60}$Fe flux, i.e.\ the signal's extension in time, still remained unanswered. 
Here, we summarize recently published results on the worldwide detection of $^{60}$Fe \cite{wallner16} and its link to the formation of the Local Superbubble (LB) \cite{breitschwerdt16}.

\section{Global traces of $^\mathbf{60}$Fe}

\begin{figure}[tbh]
\centering
\includegraphics[width=0.55\textwidth]{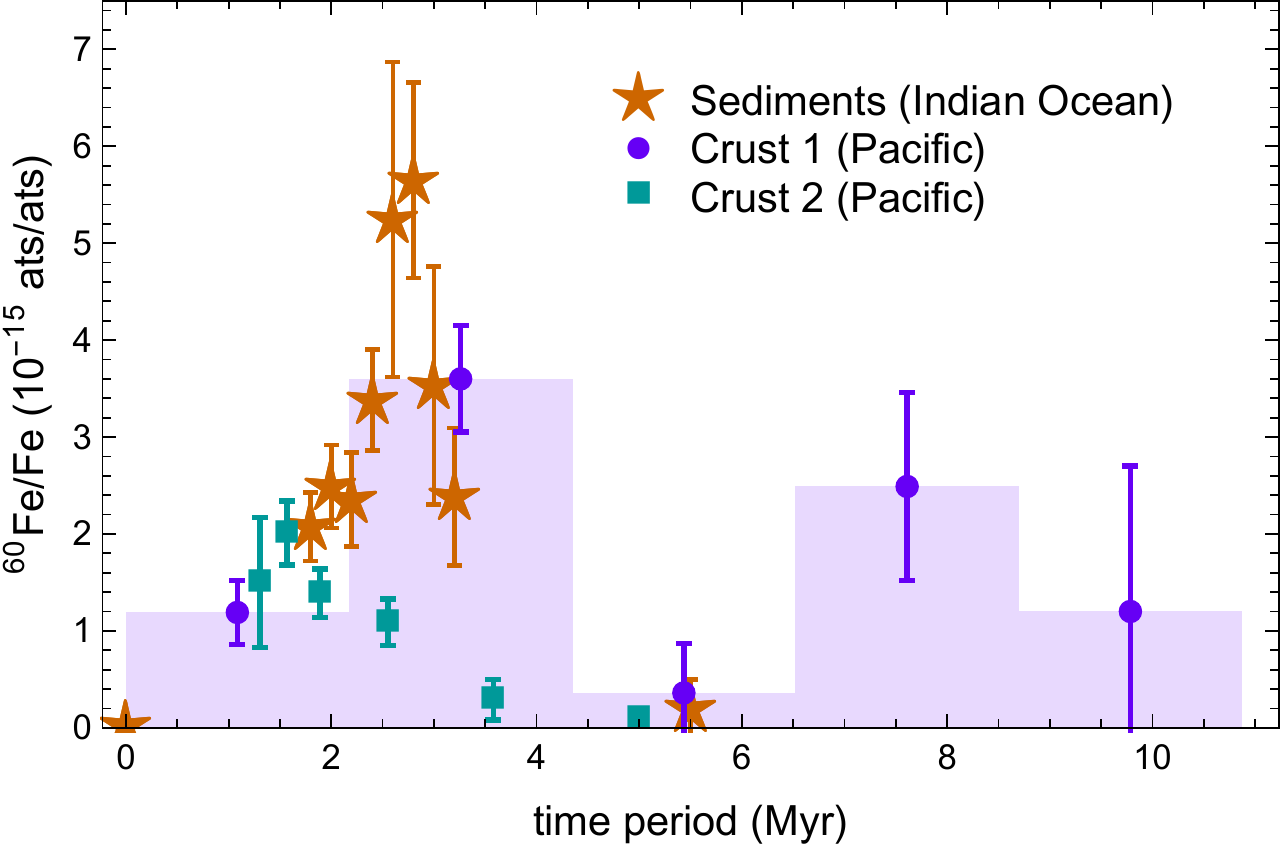}
\caption{\small Decay-corrected $^{60}$Fe/Fe versus time \cite{wallner16}. The sediment data is shown as 200\,kyr averages. Two distinct $^{60}$Fe-peaks were detected in the time-ranges of $\sim$1.5-3.5\,Myr and 6.5-8.7\,Myr in the deep-sea archives. The absolute uncertainties for the age of the samples is between 0.1\,Myr (sediments) and up to 0.5\,Myr (Crust 2), the measured $^{60}$Fe/Fe background is (4.2$\pm$1.5)$\times$10$^{-17}$.}
\label{f1}
\end{figure}

We analyzed samples from three major oceans: four deep-sea sediments from the Indian Ocean, two manganese crusts from the Pacific and two manganese nodules from the South Atlantic Ocean \cite{wallner16}. Of the slow-accumulating sediments (3-4\,mm~kyr$^{-1}$) 3 g each of 85 samples were prepared chemically \cite{feige13}, of which about 50 targets (a few mg of Fe$_2$O$_3$ each) were measured, complemented by 31 crust and 8 nodule samples. 
The cosmogenic nuclide $^{10}$Be was extracted from the samples and used for dating; the sediments and nodules were independently analyzed for $^{26}$Al. The concentrations of $^{60}$Fe in all deep-sea archives were determined with accelerator mass spectrometry (AMS) at the Heavy Ion Accelerator Facility (HIAF) at the ANU in Canberra, Australia (Fig.~\ref{f1}).

As discussed previously \cite{wallner16}, traces of $^{60}$Fe were present in the full time period covered by the deep-sea sediment of 1.7-3.2\,Myr with the exception of younger surface samples (present age) and older samples ($\sim$5\,Myr). The signal could also be verified in the two manganese nodules covering a time range up to 5.4\,Myr and in the two manganese crusts (Fig.~\ref{f1}). Interestingly, a second, older $^{60}$Fe-signal was revealed in one of the manganese crusts. With an age of 6.5-8.7\,Myr corresponding to a 5\,mm thick crustal layer this event coincides with a $^3$He-peak detected in sediments that may be associated with an asteroidal breakup-event \cite{farley06}.
This new data demonstrates a worldwide $^{60}$Fe-presence at approximately 2-3\,Myr, which has been also confirmed in magnetotactic bacteria from Pacific deep-sea sediments \cite{ludwig16}. With the iron-isotope being detected in lunar samples \cite{fimiani16}, another solar system body showed traces of recent $^{60}$Fe-deposition. 

The origin of the signal has been heavily debated. Micrometeorites were suggested as a possible source of enhanced $^{60}$Fe-flux \cite{stulee12}, however, the amount of $^{60}$Fe in the samples is less compatible with the Ni concentration, which is the target element for the production of $^{60}$Fe via cosmic rays \cite{fitoussi08, wallner16}. The evidence tends to prefer a supernova-origin consistent with discrepancies in cosmic-ray spectra \cite{kachelriess16}, with direct detection of cosmic-ray $^{60}$Fe \cite{binns16} and with the existence of the LB.

\section{The Link to the Local Bubble}

The distance of a nearby supernova to produce such a signal on Earth has been estimated between 40 and 130\,pc \cite{knie04,fry15}. We have investigated \cite{breitschwerdt16} the evolution of the solar environment and linked the formation of our LB, a region of thin hot gas embedding our solar system, to the $^{60}$Fe-peak detected in 2004 in a pacific crust \cite{knie04}, which was the only data available at the time the study was performed. The most likely sources to produce an extended structure like the LB (200\,pc in the plane, 600\,pc towards the north pole of our galaxy) are multiple stellar explosions. We have traced back in time the trajectories of a young moving group that passed the solar neighborhood and whose surviving members are now in the Scorpius-Centaurus association \cite{fuchs06}. By analyzing the observed stellar positions and velocities and their uncertainties we calculated the most probable explosion sites of the perished stars. Using an initial mass function with a variable-sized binning, we placed one star in each mass interval, thereby determining the masses of the exploded members. With a mass-age relation for main sequence lifetimes and the assumption that all stars were born at the same time, the cluster age was inferred from isochrone fitting \cite{fuchs06}, thus leading to the individual explosion times of all massive stars.

\begin{figure}[tbh]
\subfigure{\includegraphics[height =5.5cm]{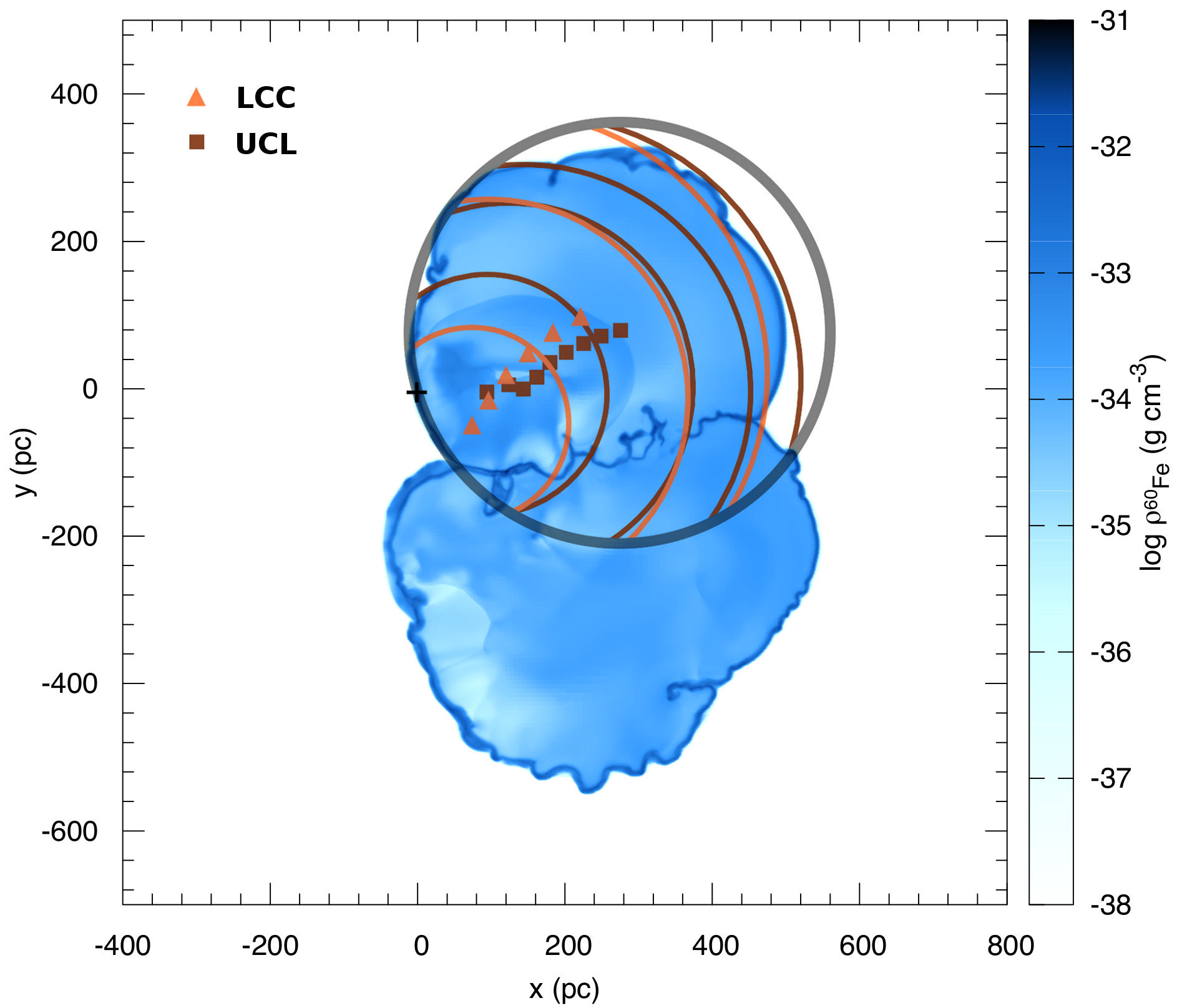}} \hspace{0.5cm}
\subfigure{\includegraphics[height =5.4cm]{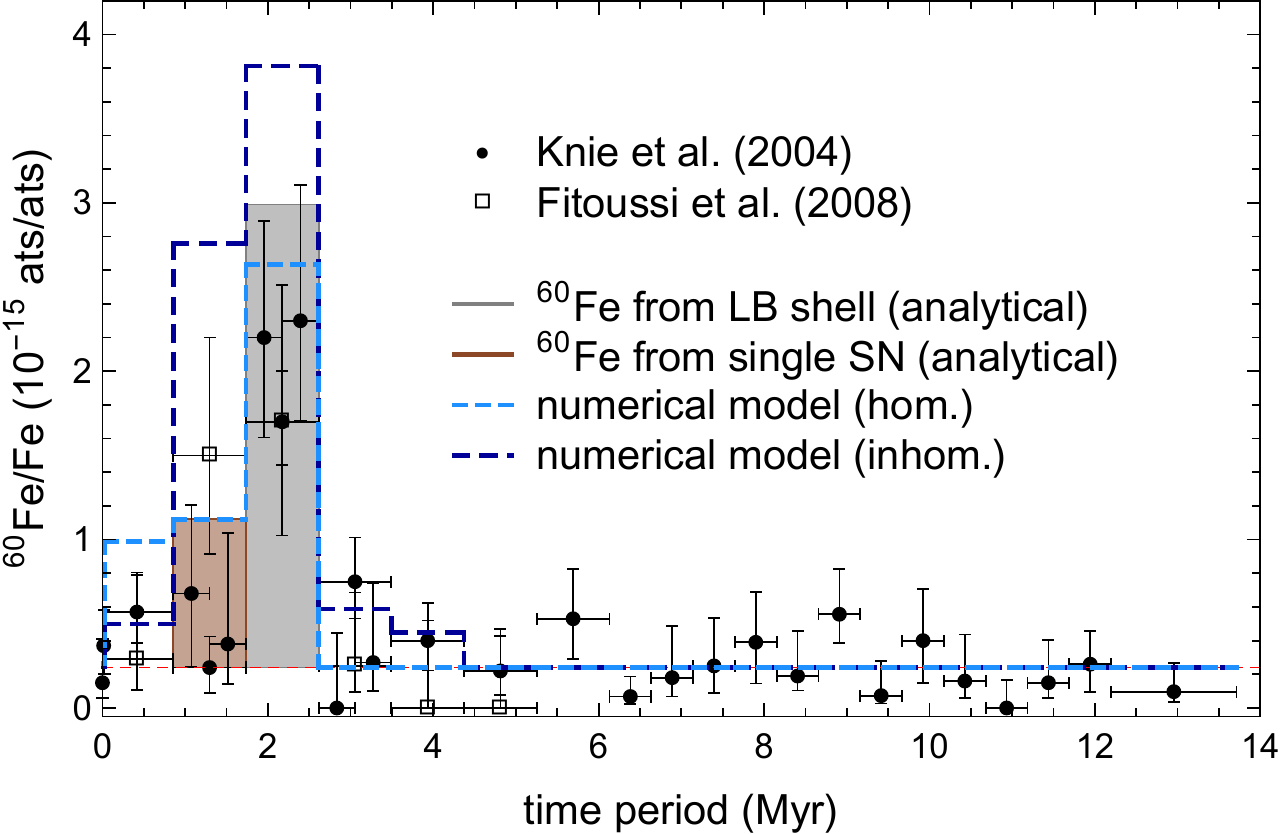}}
\caption{\small Left: Snapshot of the galactic plane 2.2\,Myr ago. Two models describe the formation of the LB using a homogeneous ambient medium with a density of 0.3\,ats~cm$^{-3}$. Analytical model: The outer shell of the LB is depicted as a gray circle, with individual remnants from single SNe expanding inside until they reach the outer shell. The surviving members of the exploded stars now belong to two subgroups of the Scorpius-Centaurus association, Lower Centaurus Crux (LCC, explosion sites as orange triangles) and Upper Centaurus Lupus (brown squares). Numerical model (blue): $^{60}$Fe density distribution in the LB (top bubble) and our neighboring superbubble Loop I. Darker regions indicate a higher density than brighter structures, the highest amount of $^{60}$Fe is distributed in the outer boundaries. In both models, the supershell of the LB sweeps over the solar system at position (0,0) $\sim$2 Myr ago. Right: $^{60}$Fe deposited in the ferromanganese crust. The red, dashed line indicates the background of the AMS measurements. The numerical and analytical models reproduce the measured $^{60}$Fe/Fe ratios \cite{knie04,fitoussi08}.}
\label{f2}
\end{figure}

Analytical and numerical methods (with homogeneous and inhomogeneous ambient medium) were used to generate the LB (described in \cite{breitschwerdt16}), as well as its neighboring Loop I Superbubble (Fig.~\ref{f2}, left). Due to the counterpressure of the surrounding medium and the Loop I, the numerical LB remains smaller than in the analytical model that neglects external forces. The $^{60}$Fe-yields ejected during the LB-forming SNe were taken from several nucleosynthesis models (see \cite{breitschwerdt16}). By calculating the fluence of $^{60}$Fe -- the number of atoms that reached the crust per cm$^2$ -- the link between the LB evolution and the $^{60}$Fe-signature could be established. Fig.~\ref{f2} (right) shows the final distribution of the isotope in the crust after exponential decay and compares it to the measured data. In both the analytical and numerical models with expansion into a homogeneous medium, it is the outer LB shell that carries $^{60}$Fe onto Earth at $\sim$2 Myr ago. This shell includes debris from 15 SNe, the first exploding 12.6\,Myr ago. Subsequently, a remnant of a 16$^\mathrm{th}$ single SN sweeps over the solar system 1.4\,Myr ago depositing $^{60}$Fe.
Due to the short lifetime of $^{60}$Fe, only the youngest three SNe exploding at distances between 91 an 106\,pc contribute significantly (62\,\%) to the terrestrial signature. In the inhomogeneous case, the LB shell arrives at 3.5\,Myr ago with three SNe following at later times producing the signal. 

\section{Outlook}

The simulations are able to link multiple nearby SN explosions that formed the LB to the measured $^{60}$Fe-signature. However, so far we have reproduced the 2-3\,Myr old crustal signal published in 2004 \cite{knie04}. In future studies we will constrain our model further with the better resolved $^{60}$Fe-signal in the deep-sea sediments from the Indian Ocean and with the 6.5-8.7\,Myr old signature. Another long-lived SN-radionuclide, $^{26}$Al, has also been measured in the deep-sea sediments. Due to the large cosmogenic atmospheric production, the detection of a SN-contribution remains challenging. Parts of the data are already published \cite{feige13}, a detailed analysis of the whole data set and its comparability to the $^{60}$Fe-signal is in preparation.

\section*{Acknowledgements}
This work was funded by the Austrian Science Fund (FWF), Project P20434 and I428 (funded via the ESF), the Australian Research Council (ARC), project no. DP14100136, the Japan Society for the Promotion of Science (JSPS) KAKENHI Grant Number 26800161, the DFG priority program 1573 and the Abschlussstipendium 2014 of the University of Vienna. 
We thank the Antarctic Marine Geology Research Facility, Florida State University, for providing samples. JF thanks the DAAD (Deutscher Akademischer Austauschdienst) for travel funding to the NIC XIV.

\end{document}